\documentclass[twocolumn,prb,aps,showpacs,superscriptaddress]{revtex4}
\usepackage{graphicx}% Include figure files
\usepackage{epstopdf}
\usepackage{dcolumn}
\usepackage{bm}
\usepackage{epsfig}
\usepackage{pictex}
\usepackage{color}
\begin{document}
\def \Nax{Na$_x$CoO$_2$~}
\def \Naxn{Na$_x$CoO$_2$}
\def \Na{Na$_{0.5}$CoO$_2$~}
\def \Nan{Na$_{0.5}$CoO$_2$}
\def \K{K$_{0.5}$CoO$_2$~}
\def \Kn{K$_{0.5}$CoO$_2$}
\def \Rb{Rb$_{0.5}$CoO$_2$~}
\def \Rbn{Rb$_{0.5}$CoO$_2$}
\newcommand {\ibid}{{\it ibid}. }
\newcommand {\etal}{{\it et al}. }
\newcommand {\etaln}{{\it et al}.}
\newcommand {\etalc}{{\it et al}., }
\title{Electronic and magnetic properties of the ionic Hubbard model
on the striped triangular lattice at 3/4 filling}
 %a strongly correlated model for the layered cobaltate: Na$_{0.5}$CoO$_2$}
\author{Jaime Merino}
\affiliation{Departamento de F\'isica Te\'orica de la Materia
Condensada,
Universidad Aut\'onoma de Madrid, Madrid 28049, Spain}
\author{Ross H. McKenzie}
\author{B. J. Powell}
\affiliation{
Centre for Organic Photonics and Electronics, School of Mathematics and Physics,
The University of Queensland,
 Brisbane 4072, Australia}
\date{\today}
\begin{abstract}
%We analyse the electronic properties of an
% ionic Hubbard model at $3/4$-filling
%on a triangular lattice with a stripe potential.
%The  on-site potential, $\Delta$, alternates on
%horizontal rows of the lattice and
%produces charge disproportionation between sites in alternating rows.
%Other parameters in the model are the intersite hopping $t$ and the
%on-site Coulomb repulsion, $U$.
We report a detailed study of a model Hamiltonian which exhibits a
rich interplay of geometrical spin frustration, strong electronic
correlations, and charge ordering. The character of the insulating
phase depends on the magnitude of $\Delta/|t|$ and on the sign of
$t$. We find a Mott insulator for $\Delta \gg U \gg |t|$; a charge
transfer insulator for $ U \gg \Delta \gg |t|$; and a correlated
covalent insulator for  $ U \gg \Delta \sim |t|$. The charge
transfer insulating state is investigated using a strong coupling
expansion. The frustration of the triangular lattice can
 lead to antiferromagnetism or ferromagnetism
depending on the sign of the hopping matrix element, $t$.
 We identify the "ring" exchange
process around a triangular plaquette which determines the
sign of
the magnetic  interactions.
 Exact diagonalization calculations are performed on
the model for a wide range of parameters and compared to
the strong coupling expansion.
The regime $ U \gg \Delta \sim |t|$ and $t<0$ is
relevant to \Nan.
The calculated optical
conductivity and the spectral density are discussed in the light of recent experiments on \Nan.
\end{abstract}
% 71.10.Fd  Lattice fermion models (Hubbard model, etc.)
% 71.15.-m  Methods of electronic structure calculations
% 71.27.+a  Strongly correlated electron systems; heavy fermions
\maketitle
\section{Introduction}
Many strongly correlated electron materials  exhibit a subtle
competition between different magnetic and charge ordered states,
and between metallic, insulating, and superconducting phases. Widely
studied (and poorly understood) materials include cuprate
superconductors,\cite{lee} organic charge transfer
salts,\cite{ishiguro} manganites with colossal
magnetoresistance,\cite{CMR} heavy fermion compounds,\cite{si} and
the  iron  pnicitide superconductors.\cite{chen} A fundamental
theoretical challenge is explaining the hierarchy of energy scales
and competing phases in these materials. The energy scales (such as
the bandwidth and Coulomb repulsion) associated with the relevant
electronic orbitals (and microscopic Hamiltonians such as Hubbard
models) are typically of the order of eV. In contrast, the energy
scales associated with the temperature and magnetic field
dependences of transport properties and energy differences between
competing phases are often  several orders of magnitude smaller.
Frustration of spin or charge ordering by competing interactions due
to the geometry of the crystal lattice can enhance these effects. In
addition, it is not clear what physical changes are produced by
chemical doping. For example, does adding charge carriers just
change the band filling or are there significant effects due to the
associated disorder and changes in the electronic structure?

Here we report a detailed study of a specific strongly correlated
electron model, the ionic Hubbard model on the triangular lattice at
3/4 filling with a stripe potential. The model illustrates how the
interplay of geometric frustration and strong correlations lead to
competition between different magnetic orders, charge ordering,
metallic, and insulating behaviours. One concrete realisation of the
model is that it may be the simplest many-body Hamiltonian that can
describe Na$_{0.5}$CoO$_2$.\cite{MPM,MPM1,PMM} Elsewhere we have
reviewed experimental results on this material and described recent
theoretical attempts to describe its unusual properties.\cite{PMM}
When the filling $x$ in \Nax is close to other commensurate values,
such as $1/3, 2/3,$ or $3/4$, the system is still described by an
ionic Hubbard model but the on-site potential has a different form
and commensurability,
 depending on the ordering arrangement of the sodium ions.\cite{MPM}
At incommensurate values of $x$ one expects phase coexistence of
multiple Na-ordering phases.\cite{Zhang}

The rest of the paper is organized as follows.  In Sec.
\ref{sec:model} we introduce an ionic Hubbard model on a triangular
lattice including a discussion of its phase diagram. In Sec.
\ref{sec:lanczos} we analyse the model's ground state properties
using the Lanczos exact diagonalization technique on finite-size
clusters. Dynamical properties such as the spectral density and
optical conductivity are discussed in Sec. \ref{sec:dynamic}.
Finally, a summary of the main results and their relevance to \Na is
given in Sec. \ref{sec:concl}. We have  also studied the same model
using a complementary method, mean-field slave bosons.\cite{PMM} At
appropriate places in the paper we compare and contrast the results.

\section{The Ionic Hubbard Model on a triangular lattice}
\label{sec:model}

\begin{figure}
\begin{center}
\epsfig{file=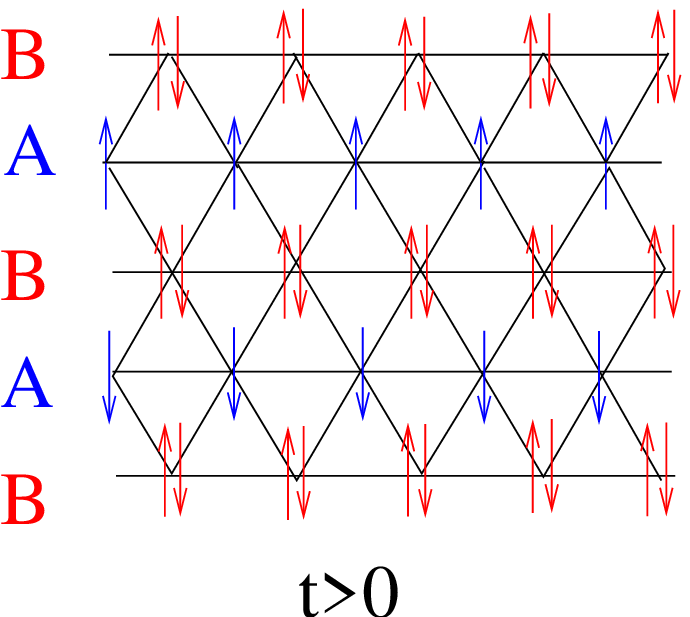,width=4.2cm,clip=}
\epsfig{file=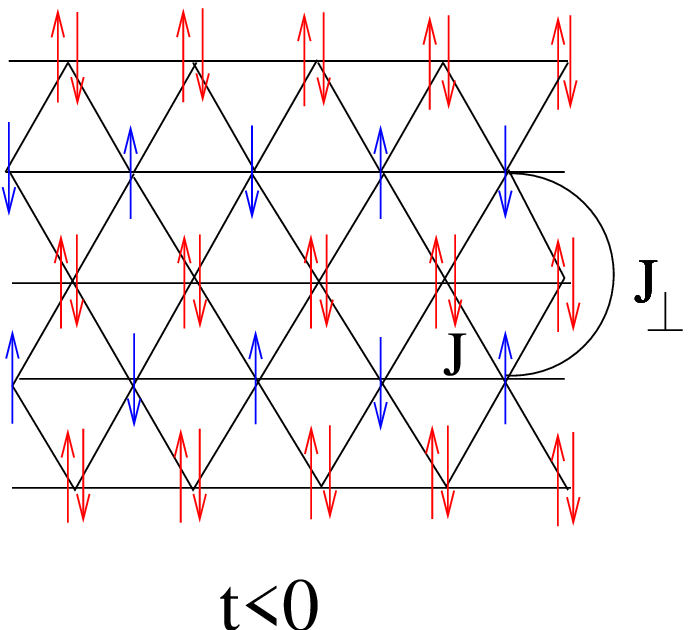,width=4.2cm,clip=}
\caption{\label{figstripes}(Color online)
Spin and charge order in the $3/4$-filled Ionic Hubbard model (\ref{ham}) on
a triangular lattice in the limit $U>>\Delta>>|t|$. $A$ and $B$ denote the
inequivalent sites of the lattice.
%The charge and
%spin pattern for $t<0$ is consistent with observations on \Nan.
C-type antiferromagnetism (left) is found for $t>0$ in contrast to 
G-type antferromagnetism (right) for $t<0$.
The exchange couplings $J$ and $J_\perp$ are defined
in Eq. (\ref{superex}) for the appropriate $t-J$-model in Eq. (\ref{tJ}). 
A ferromagnetic 
exchange coupling, $J$, between neighboring A-sites 
occurs for the parameter range: $0 < 5t<\Delta <\sqrt{2U}$.}
\end{center}
\end{figure}

The Hamiltonian of the ionic Hubbard model is
\begin{equation}
H=-t\sum_{<ij>\sigma} (c^+_{i\sigma} c_{j\sigma}+ c^+_{j\sigma}
c_{i\sigma}) +U \sum_i n_{i \uparrow} n_{i \downarrow}
+\sum_{i\sigma} \epsilon_i n_{i \sigma}, \label{ham}
\end{equation}
where $c^+_{i\sigma}$ creates an electron with spin $\sigma$ at site $i$,
 $t$ is the hopping amplitude between neighbouring sites and
$U$ is the effective on-site Coulomb repulsion energy between two
electrons. We set $\epsilon_i=\Delta/2 $ for the A-sites 
and $\epsilon_i=-\Delta/2$ for B-sites (cf. Fig. \ref{figstripes}). The
A-sites form rows which alternate with the B-sites of the
triangular lattice. By a particle-hole transformation:
$c^+_{i\sigma} \rightarrow h_{i\sigma} $ model (\ref{ham}) becomes a
$1/4$-filled (with holes) ionic Hubbard model with the sign
transformation: $t \rightarrow -t$ and $\Delta \rightarrow -\Delta$.

In what follows we will discuss 
 the charge gap which is defined for the
model on a finite lattice with $N$ electrons and $2N/3$ lattice
sites by
\begin{equation}
\Delta_c \equiv E_0(N+1)+E_0(N-1)-2E_0(N),
\label{chargegap0}
\end{equation}
where $E_0(M)$ is the ground state energy of the system
with $M$ electrons.

\subsection{Previous theories of the ionic Hubbard model}

The ionic Hubbard model (\ref{ham}) on the striped
triangular lattice of Fig. \ref{figstripes} considered here 
contains geometrical frustration. This is in contrast to most 
previous work which has focused on bipartite, i.e., unfrustrated, lattices
with different site energies, $\Delta$, on each of the bipartite
sublattices and at half-filling. Some of the interest in this model
can be appreciated from the half-filled atomic limit ($t=0$). For
$U>\Delta$ the charge gap for the addition of particles defined in Eq. (\ref{chargegap0}), 
$\Delta_c=U-\Delta$; thus the system is a Mott insulator. But, for $U<\Delta$,
$\Delta_c=\Delta-U$; and the system is a band insulator. At the
point $U=\Delta$ this gap vanishes. Therefore a key question is what
happens at the band to Mott insulator transition away from the
atomic limit; in particular what happens to the gapless point - does
it expand give a metallic phase? Further impetus comes from the
proposals that the ionic Hubbard model is important for
understanding ferroelectric perovskites, \cite{Egami} organic charge
transfer salts,\cite{Nagaosa,incomen} transition metal oxide
heterostructures,\cite{Millis} and non-linear electronic
polarizability in transition metal oxides,\cite{Ishihara} and, as
discussed below the rich electronic phases observed in A$_x$CoO$_2$
[A=Na, K, Rb] (Ref.  \onlinecite{PMM}).

The most studied case is the half filled one-dimensional chain with
different site energies for odd and even numbered sites. This model
shows three distinct insulating phases: a band insulator; a
(ferroelectric) bond order wave insulator; and a Mott
insulator.\cite{Kampf} Metallic behaviour appears to be limited to
the point in the phase diagram where band insulator gives way to the
bond order wave insulator.\cite{Kampf} Continuum limit bosonization
calculations suggest that adding a next nearest neighbour hopping,
$t'$, (which is equivalent to studying the zigzag chain) induces a
large metallic region in the phase diagram;\cite{Japaridze}
suggesting that even in one-dimension frustration already plays an
important role in the ionic Hubbard model.

The infinite-dimensional ionic Hubbard model has been studied using
dynamical mean-field theory (DMFT),\cite{Garg,Craco}
which treats the on-site quantum dynamics exactly
but ignores spatial correlations such
as those associated with antiferromagnetic exchange.
  These papers studied the, bipartite,
Bethe lattice and
found that a metallic phase separates the band insulating phase from
the Mott insulating phases, in, at least, some parts of the phase
diagram.

In two-dimensions most previous work has focused on the half-filled
square lattice with site energies alternating in a checkerboard
pattern. This model has been studied using both cluster
DMFT\cite{Kancharla} and determinant quantum Monte Carlo
(DQMC).\cite{ParisPRL,Bouadim} These studies all suggested that a
phase with non-zero spectral weight at the Fermi energy exists
between the band insulator and Mott insulator phases, at least in
some of the phase diagram. However, the has been some debate over
whether this phase is metallic\cite{ParisPRL,Bouadim} or bond
ordered.\cite{Kancharla}

There has been far less work on the ionic Hubbard model away from
half-filling. However, Penc \etal\cite{Penc} studied the
quarter-filled ionic Hubbard model on the zigzag ladder and found a
competition between ferromagnetism and a paramagnetic phase with
strong antiferromagnetic correlations. Bouadim \etaln,\cite{Bouadim}
studied the ionic Hubbard model on a square lattice with a
checkerboard potential across all possible fillings with DQMC. The
most interesting features they found, away from half filling, were
Mott insulation at quarter and three quarters filling (which are
related by the particle-hole symmetry of this model). Bouadim \etal
did not find any evidence of magnetic order in this phase. However,
as the magnetic interactions are ${\cal O}(t^4/U\Delta^2)$ the
absence of magnetic order may be due to non-zero value of the
temperatures they studied.

As well as bipartite arrangements of the different site energies there has also been considerable interest
in random arrangements
of site energies.
 Laad \etal\cite{Laad} studied a system with a gaussian density of states and a bimodal distribution of
site energies  in infinite
dimensions for various impurity concentrations, $n$, and fillings, $1-\delta$. They found insulating
states for $\delta=1-n$ and
sufficiently large $\Delta$ and $U$.   Byczuk \etal have used DMFT to study the (frustrated)
fcc\cite{ByczukPRL} and (bipartite)
Bethe\cite{ByczukPRB} lattices in infinite dimensions with
a bimodal distribution of site energies and half the sites taking each value of the site energy. At
one-quarter filling they found that
metal insulator transitions occur on both
 lattices when both $U$ and $\Delta$ are sufficiently large. They also noted that DMFT does not capture
some of the possible effects of the
disorder, such as Anderson insulating phases.
  In two dimensions Paris \etal\cite{ParisPRL} used DQMC to study the
square lattice at a range of fillings with 1/8 of the sites randomly
chosen to have a different site energy than the rest of the lattice.
They found that this model displayed Mott insulating, band
insulating, Anderson insulating, and metallic phases.

Marianetti and Kotliar\cite{Kotliar} simplified our suggestion\cite{MPM} of
that Eq. (\ref{ham}) is the appropriate effective low Hamiltonian for \Nax by further assuming that
Na-ordering is of secondary
importance and hence treated the potential due to the Na ions as random. They then used density functional
theory to show that the distribution of Co site
 energies is bimodal and to parameterise the Hamiltonain (\ref{ham})
 for $x=0.3$ and $0.7$. Finally they calculated the high temperature
($T\gtrsim100$ K) susceptibility for these dopings and found them to
be in qualitative agreement with experiment. We deal here with the
case $x=0.5$ for which Na-ordering of the stripe-type (cf. Fig
\ref{figstripes}) has been observed in
experiments.\cite{Na-ordering,Foo}

\subsection{The non-interacting model ($U=0$)}

For $U=0$ model (\ref{ham}) can be diagonalized straightforwardly
leading to two bands, denoted $\pm$. We introduce creation and
destruction operators:
\begin{equation}
c^\dagger_{{\bf k}\pm\sigma}=\alpha_{{\bf k}\mp}(c^+_{A{\bf k}\sigma}+
A_{{\bf k}\mp}c^+_{B{\bf k}\sigma}) ,
\label{norm}
\end{equation}
where $c^+_{A{\bf k}\sigma}$ and $c^+_{B{\bf k}\sigma}$ act on the Bloch states
associated with the $A$ and $B$ sublattices, respectively, and
\begin{equation}
A_{{\bf k}\pm}={ \Delta/2 \pm \sqrt{\Delta^2/4+(4t\cos(k_x/2)\cos(k_y\sqrt{3}/2))^2}
\over 4t \cos(k_x/2) \cos(k_y\sqrt{3}/2)  },
\label{mixing}
\end{equation}
with the normalization constant: $\alpha_{{\bf k}\pm}=1/\sqrt{1+|A_{{\bf k}\pm}|^2}$.
The energy dispersion of the two bands is:
\begin{equation}
\epsilon_{\pm}({\bf k})=-2t\cos{k_x}\pm\sqrt{\Delta^2/4+(4t\cos(k_x/2) \cos(k_y \sqrt{3}/2))^2},
\label{tb}
\end{equation}
with $k_x$ and $k_y$ defined in the reduced $(1\times \sqrt{3})$
Brillouin zone with lattice parameter $a = 1$. At $3/4$-filling and
for any $\Delta$, there is always at least one band
crossing the Fermi energy and so the system is metallic. The $+$ band
is half-filled and the $-$ band filled for $t>0$ and $\Delta>0$, whereas for $t<0$
this only occurs for $\Delta>0.64|t|$.
 For $\Delta=0$ there is only one
 band, which has  a width of $W=9|t|$.

\begin{figure}
\begin{center}
\epsfig{file=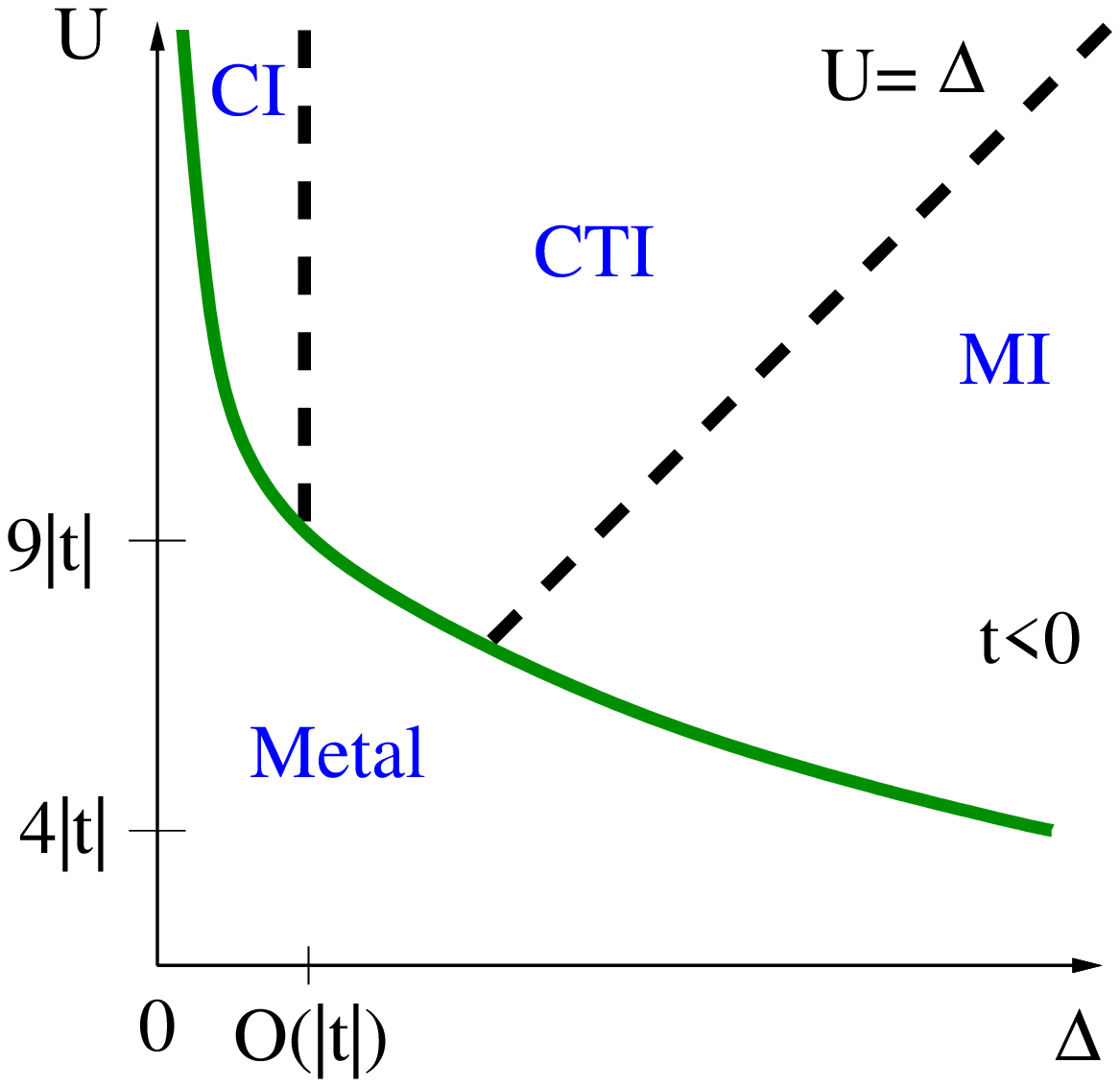,width=5.5cm,clip=}
\epsfig{file=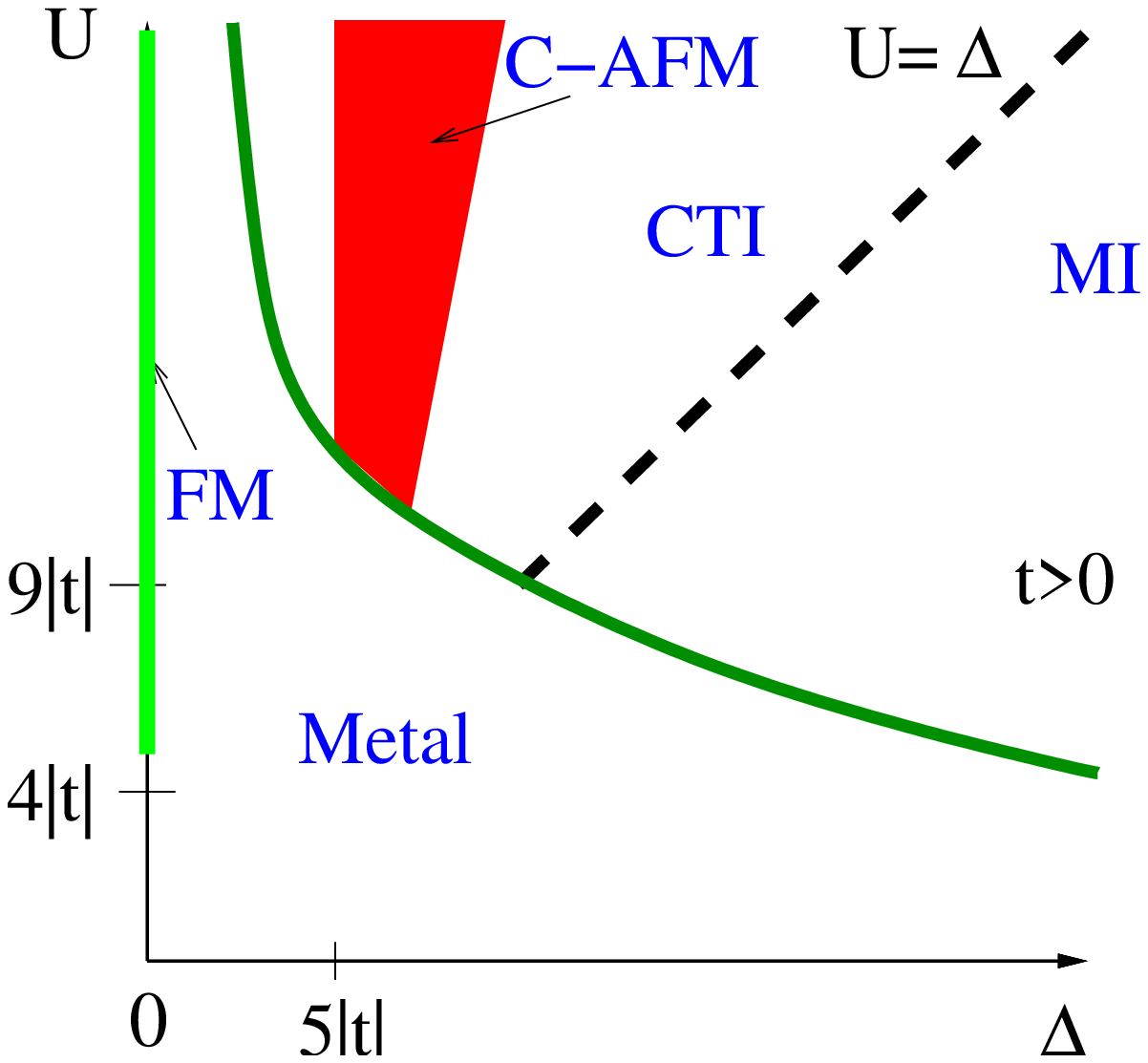,width=5.5cm,clip=}
\caption{\label{figphased}(Color online)
Schematic phase diagram of the Ionic Hubbard model \ref{ham} on
a striped triangular lattice at $3/4$-filling. The
transition lines are based on the lowest order corrections in a strong coupling
analysis and on Lanczos diagonalization calculations. The $t<0$ (top panel)
case is relevant to the Na$_{0.5}$CoO$_2$ insulator.
Insulating phases at strong coupling, $U>W$, of different types are found ranging from
a charge transfer insulator (CTI),
a Mott Insulator (MI) and a covalent insulator (CI). The bandwidth of the
model for $\Delta=0$ (the isotropic triangular lattice), $W=9|t|$, is effectively 
reduced to $4|t|$ corresponding to one-dimensional chains as $\Delta$ increases. 
At exactly  $\Delta=\infty$, the system is insulating for any nonzero $U$ as expected
for a half-filled Hubbard chain due to Umklapp processes.
The blue line is an estimate of the critical $U$ for
the metal-to-insulator transition which follows the effective bandwidth dependence with $\Delta$.
Insulating phases for $t<0$ display G-type antiferromagnetic (AFM) (see Fig. \ref{figstripes})
correlations whereas a C-AFM region (see Fig. \ref{figstripes}) for $t>0$
is obtained from the condition, $J<0$, to Equation (\ref{superex}) which we assume
valid for $U \gtrsim 9|t|$. The marked $\Delta=0$ axis for $t>0$ and above $U\approx 5|t|$ 
indicates the occurrence of ferromagnetism as predicted by DMFT of the Hubbard model on a 
isotropic triangular lattice \cite{MPM}.
}
\end{center}
\end{figure}

\subsection{Phase diagram}

The ionic Hubbard model (\ref{ham}) on a triangular lattice contains
a rich phase diagram resulting from the interplay between
geometrical frustration, strong Coulomb repulsion and
charge ordering phenomena. A schematic
phase diagram can be constructed
by first considering some simple limits:

{\it (i) $t=0$ (atomic limit)}:  all A sites are singly occupied while B sites are doubly
occupied. The system is insulating with a charge gap: $\Delta_c=\text{min}(\Delta,U)$. For
$\Delta>U$ it is a Mott insulator (MI) with $\Delta_c=U$, whereas for $\Delta<U$ it is a
charge transfer insulator (CTI) with $\Delta_c=\Delta$.

{\it (ii) $U=0$ (non-interacting limit):}
As discussed above the model is always metallic
regardless the value of $\Delta$.

{\it (iii)  $\Delta=0$:} For any $U$ the
model reduces to the regular  Hubbard model on
the isotropic triangular lattice at 3/4 filling.
For large $U/|t|$, it is equivalent to the $t-J$ model
on the triangular lattice.
Dynamical mean-field theory calculations\cite{MPM,giamarchi}
give a ground state that is metallic.
DMFT\cite{MPM} calculations on the Hubbard model for large $U$
and variational Monte Carlo calculations on the $t-J$ model
give this metallic ground state as paramagnetic (ferromagnetic)
for $t<0$ ($t>0$).\cite{weber}
%Exact diagonalisation  studies of the $t-J$ model
%at 3/4 filling also suggest
%that the ground state is metallic.\cite{haerter,haerter2}

{\it (iv) $\Delta=\infty$:} As the B-sites can be completely projected out from the Hilbert space,
the model
is mapped onto decoupled half-filled Hubbard chains.
 Hence, the system is (Mott) insulating for any non-zero
positive $U$; i.e., there is a charge gap
$\Delta_c\ne 0$, and there are antiferromagnetic correlations
(with power law decay) and  no spin gap.

{\it (v) $U>>\Delta>>|t| \ne 0$:}
For finite
but small $t$, virtual hopping processes lead to effective magnetic
exchange couplings between the $A$ sites. The effective low-energy
$t-J-J_{diag}-J_\perp$ Hamiltonian for the holes is
\begin{eqnarray}
H &=&t\sum_{ij\sigma} P(h^+_{i\sigma} h_{j\sigma}+ h^+_{j\sigma}
h_{i\sigma})P +J \sum_{\{ij\}} \left[ {\bf S_i}\cdot{\bf S_j}-{n_in_j \over 4} \right]
\nonumber \\
&+& J_{diag} \sum_{(ij)} \left[ {\bf S_i}\cdot{\bf S_j}-{n_in_j \over 4} \right]
\nonumber \\
&+& J_\perp \sum_{[ij]} \left[ {\bf S_i}\cdot{\bf S_j}-{n_in_j \over 4} \right]
-\sum_{i\sigma} \epsilon_i h^+_{i \sigma}h_{i \sigma}, \label{tJ}
\end{eqnarray}
where $\{...\}$, and $[...]$ denote sums over intra-A-chain,
inter-A-chain sites, respectively. The sum over $(...)$ is between
an A and nearest-neighbour B-sites. The projector
$P=\Pi_i\left[1-n_{i \uparrow} n_{i\downarrow} \right]$ forbids
double occupation of holes on any lattice site. The dynamics of the
electron-doped system relevant to \Nax with electron occupation
$1+x$ is related to the hole-doped system, with filling $1-x$,
through the replacement: $t \rightarrow -t$ and $\epsilon_i
\rightarrow -\epsilon_i$ leaving the exchange parameters unchanged.

The exchange couplings $J$, $J_\perp$ and $J_{diag}$ can be obtained
through a strong coupling expansion using Raleygh-Schr\"odinger
perturbation theory on the hopping term around the configuration in
which all B sites are doubly occupied and A-sites singly occupied
(see Appendix \ref{sec:appI}). This leads to an effective exchange
coupling between electrons in $A$-sites in the horizontal direction:
\begin{equation}
 J={4t^2 \over U}-{8t^3 \over \Delta^2}-{16t^3 \over \Delta U}
+{\cal O}(t^4),
\label{superex}
\end{equation}
and in the perpendicular direction
\begin{equation}
J_{\perp}= {16t^4 \over \Delta^2} \left[{1\over
U}+{1\over 2 \Delta+U}+{1 \over 2 \Delta}\right]
+{\cal O}(t^5).
\label{superexperp}
\end{equation}
The exchange coupling $J_{diag}$ between
$A$ and $B$ sites is
\begin{equation}
J_{diag}=2t^2 \left[{1 \over U+\Delta} + {1 \over U-\Delta}\right]
+{\cal O}(t^3),
\end{equation}
which is blocked if the $B$ sites are doubly occupied but recovers
the correct $4t^2/U$ exchange interaction as $\Delta \rightarrow 0$.

The second and third
terms in $J$ are antiferromagnetic (AFM) for $t<0$
and ferromagnetic (FM) for $t>0$.  Higher order contributions to $J$
% of ${\cal O}(({t}/{\Delta})^4)$
are AF and can be found in
Appendix \ref{sec:appI}.
%The magnetic exchange coupling between the A-chains in the
%perpendicular direction, $J_\perp$, is of ${\cal O}(({t}/{\Delta})^4)$.

A schematic phase diagram of the $3/4$-filled ionic Hubbard model on a
triangular lattice (\ref{ham}) is shown in Fig. \ref{figphased}.
The transition lines are extracted from the
limits {\it (i)-(v)} discussed above and exact diagonalization calculations for
intermediate parameter regimes. Apart from the
Mott insulator (MI) and charge transfer insulator (CTI),
our numerical analysis suggests the presence of a covalent insulator
(CI) in the range $\Delta \sim O(|t|)$ and $U>>|t|$.
Depending on the sign of $t$, different spin arrangements occur
as shown in Fig. \ref{figstripes}. The condition $J=0$ separates
AF from the FM region which occurs in the parameter
range: $5t<\Delta < \sqrt{2U}$ and is plotted in Fig. \ref{figphased}.

\begin{figure}
\begin{center}
\epsfig{file=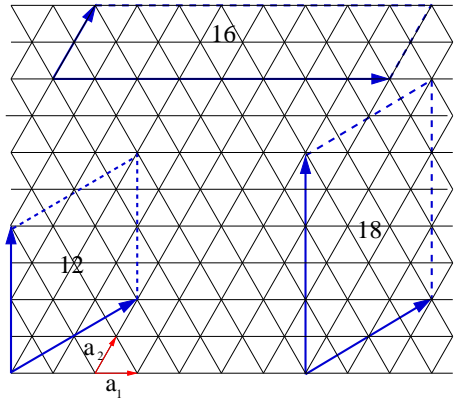,width=6.cm,clip=}
\caption{\label{figclusters}(Color online)
Cluster shapes of different sizes used in exact diagonalization calculations.
}
\end{center}
\end{figure}

\subsection{The model on two and four-site clusters}

In this section we explore the nature of the ground state of the
model (\ref{ham}) on two and four-site clusters. The two-site
cluster incorporates charge ordering phenomena in the presence of
on-site Coulomb interaction. The four-site toy model also contains
geometrical frustration effects present in the full model
(\ref{ham}). Ground state properties of the clusters are discussed
in terms of valence-bond (VB) theory \cite{pauling} when
appropriate. For $t<0$ and $U>>|t|$ the ground state wavefunction is
accurately described by the resonance between possible valence
bonds. Our analysis indicates that the charge gap of the clusters is
enhanced with $\Delta$ due to the differences between the
two-electron and three-electron bonds between the different $A$ and $B$ sites.

\subsubsection{Two-site cluster}

\begin{figure}
\begin{center}
\epsfig{file=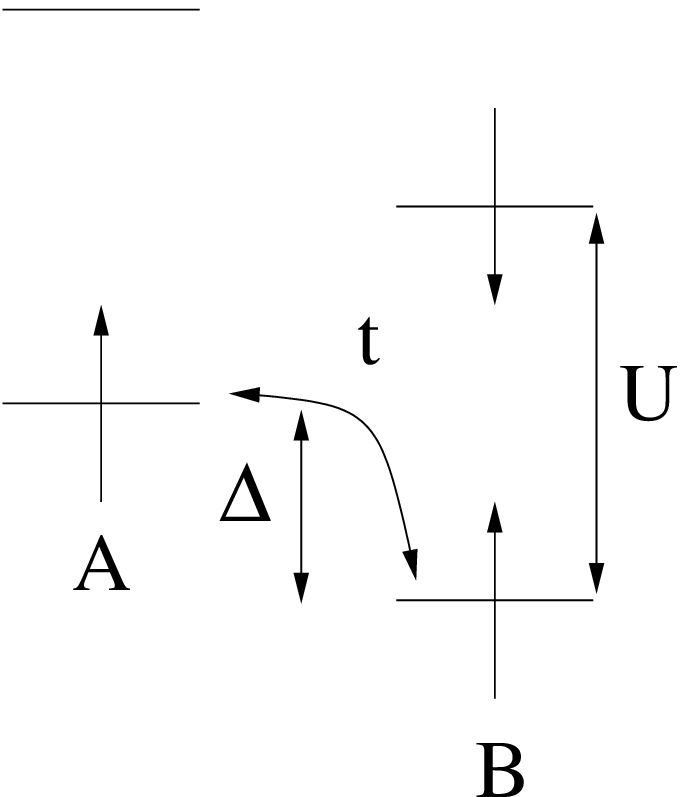,width=4.0cm,clip=}
\epsfig{file=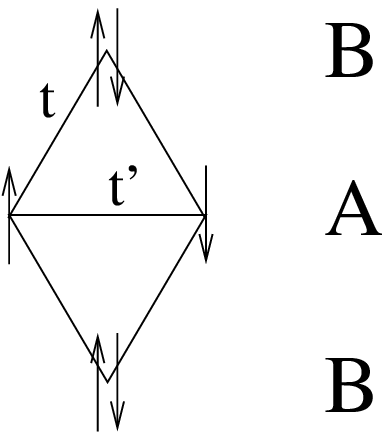,width=3.0cm,clip=}
\caption{\label{fig2sites}(Color online)
Ionic Hubbard model on two- and four-site clusters.
The energy level diagram for 3 electrons on two sites (left) and
the four-site cluster with six electrons (right).
}
\end{center}
\end{figure}

We first consider three electrons in two inequivalent sites
(one $A$, the other B) separated by an energy $\Delta$.
The energy levels for
this cluster are sketched in Fig. \ref{fig2sites}.
The Hamiltonian  is:
\begin{eqnarray}
H&=&-t (c^+_{A\sigma} c_{B\sigma}+ c^+_{B\sigma}c_{A\sigma})
+U (n_{A \uparrow} n_{A \downarrow}+n_{B \uparrow} n_{B \downarrow})
\nonumber \\
&+&\Delta/2 (n_A-n_B). \label{ham2site}
\end{eqnarray}

For $U=0$, the charge gap of the cluster is $\Delta_c=0$ for any $\Delta$ due to
the degeneracy of the ground state.
In order to evaluate the dependence of the gap on $\Delta$
 we
first obtain the ground state energies with $N=$2, 3 and 4 electrons:
\begin{eqnarray}
E_0(2) &=&-2t^2 \left[ {1 \over U+\Delta} + {1 \over U-\Delta}  \right], U>>|t|
\nonumber \\
E_0(3) &=& U-{\sqrt{\Delta^2+4t^2}\over 2}
\nonumber \\
E_0(4) &=& 2U.
\end{eqnarray}
In the limit $\Delta \rightarrow 0$ and $U>>|t|$, the charge gap of the cluster
is
\begin{equation}
\Delta_c \approx 2|t| +\Delta^2\left({1\over 4t}-{4t^2\over U^3}\right)-{4t^2 \over U}.
\end{equation}
The first contribution to $\Delta_c$ is present even for $\Delta=0$
as expected from the bonding-antibonding splitting of the cluster
and will go to zero in the infinite system. The term proportional to
$\Delta^2$ comes from the different dependances of $E(3)$ and $E(2)$
on $\Delta$: $E(2)$ has a weaker dependence than $E(3)$. This is due
to the different nature of the two-electron and the three-electron
bond. The former is accurately described by a correlated VB between
an electron on an A site and an electron on a B site whereas the
latter is described by a single hole in an antibonding ``molecular''
orbital. The two-electron and three-electron bond energies (the
energy needed to break a bond between inequivalent sites) are
$\Delta/2-\sqrt{\Delta^2+4t^2}/2$ and $E_0(2)$, respectively. Hence,
the two-electron bond becomes weaker with $\Delta$ while the
three-electron bond is strengthened with $\Delta$. This is known
from quantum chemistry \cite{pauling} and can be understood as being
a consequence of the presence or absence of Coulomb repulsion
between electrons.

\subsubsection{Four-site cluster}

We consider the four-site cluster (Fig. \ref{fig2sites})
 with two A-sites
shifted by $+\Delta/2$ and two B-sites shifted by $-\Delta/2$.
The cluster shown contains $N=6$ electrons (correponding to $3/4$-filling).
%in the atomic limit $t=0$ with $U>\Delta$.
The model Hamiltonian in this case is
\begin{eqnarray}
H &=&-t \sum_{i\in A,j\in B} (c^+_{i\sigma} c_{j\sigma}+ c^+_{j\sigma} c_{i\sigma})
-t' \sum_{i,j\in A} (c^+_{i\sigma} c_{j\sigma}+ c^+_{j\sigma} c_{i\sigma})
\nonumber \\
&+&U \sum_i n_{i\uparrow} n_{i\downarrow}
+\Delta/2 \sum_{i\in A,j \in B}(n_{i}-n_{j}). \label{ham4site}
\end{eqnarray}

\begin{figure}
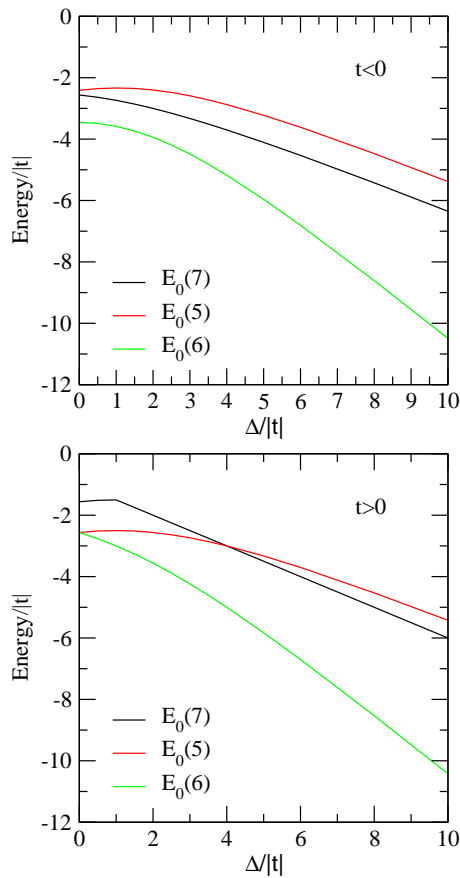

\begin{center}
\epsfig{file=fig5a.eps,width=6.0cm,clip=}
\epsfig{file=fig5b.eps,width=6.0cm,clip=}
\caption{\label{ener4sites}(Color online) Dependence of
ground state energies with $\Delta$
for the four-site cluster with $t'=t$ and $U=100|t|$. The
Coulomb interaction $U$, $2U$ and $3U$ have been substracted
from the total energies $E_0(5)$, $E_0(6)$ and $E_0(7)$, respectively,
for convenience.}
\end{center}
\end{figure}

\begin{figure}
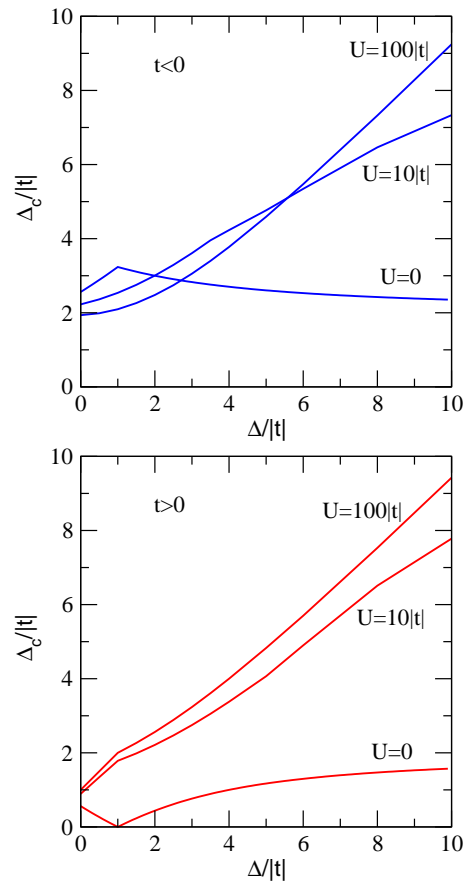

\begin{center}
\epsfig{file=fig6a.eps,width=6.0cm,clip=}
\epsfig{file=fig6b.eps,width=6.0cm,clip=}
\caption{\label{gap4sites}(Color online) Dependence of the charge gap
on   $\Delta$ for the four-site cluster. Note the increase
of $\Delta_c$ for any $\Delta$ for $U>>|t|$ of Fig. (\ref{ener4sites}) and
also how the dependence for $\Delta > |t|$ is quite different with $\Delta<|t|$.
}

\end{center}
\end{figure}

We first discuss the $\Delta \rightarrow 0$ limit. The exact ground
state energies of the fully frustrated $t'=t$ cluster for $N=5$, 6
and 7 electrons are plotted in Fig. \ref{ener4sites} for $U=100|t|$.
Consistent with the results for the two-site cluster we find that,
$E_0(5)$, has the weakest dependence on $\Delta$ of all the ground
state energies. The ground state wavefunction for $N=6$ can be well
described in terms of resonating valence bonds as shown in the
Appendix \ref{sec:appII}. In contrast, the wavefunction for $N=7$
consists of a single hole hopping around the cluster and so contains
no Coulomb interaction effects. In this case a ``molecular'' orbital
with a single hole describes the cluster and its energy, $E_0(7)$,
has the strongest dependence with $\Delta$. The different behavior
of $E_0(N)$, $E_0(N+1)$ and $E_0(N-1)$ is responsible for the
increase of the charge gap with $\Delta$ as shown in Fig.
\ref{gap4sites}. This behavior is in contrast to the $U=0$ case
plotted in the figure.

%\begin{eqnarray}
%\Delta_c &=& {3t'\over 2} - \Delta/2+ {1 \over 2}\sqrt{(\Delta+t')^2+16 t^2}, \Delta > |t|
%\nonumber \\
%\Delta_c &=& {t'\over 2} + \Delta/2+ {1 \over 2}\sqrt{(\Delta+t')^2+16 t^2}, \Delta < |t|,
%\end{eqnarray}
%At $\Delta=|t|$ there is a crossing of the one-electron levels. The
%gap for the non-interacting cluster is plotted in Fig. \ref{gap4sites}.
%Note that the gap goes to zero as $\Delta \rightarrow \infty$.

{\it Effects of frustration.} We now discuss the four-site cluster
with $t'=0$. In this case analytical formulas may be obtained for
$U=\infty$. The sign of the hopping is irrelevant here in contrast
to the $t' \ne t$ case.
%Within the $6$-particle sector we find that the energy spectra
%is: $-\sqrt{8t^2+\Delta^2}$,0 and $\sqrt{8t^2+\Delta^2}$.
The ground state energies are:
\begin{eqnarray}
E_0(6) &=& -\sqrt{8t^2+\Delta^2}
\nonumber\\
E_0(7) &=& -t-{\sqrt{4t^2+\Delta^2}\over 2} .
\end{eqnarray}
% and for
%$\Delta>>t$,  $E_0(6) \approx -\Delta-4t^2/\Delta$, which gains
%energy due to hybridization between the A and B sites.
%For $7$-particles the energy level structure is given by:
%$-t-\sqrt{4t^2+\Delta^2}/2, -t+\sqrt{4t^2+\Delta^2}/2,
%t-\sqrt{4t^2+\Delta^2}/2, t+\sqrt{4t^2+\Delta^2}/2$.
%In this case the energy in the $\Delta >>t$ limit can
%be approximated through: $E_0(7)=-t-\Delta/2-t^2/\Delta$.
In the limit $\Delta \rightarrow 0$, we find
$E_0(7) \approx -2t-{\Delta^2 \over 4|t|}$ and
$E_0(6) \approx -2\sqrt{2}t-{\sqrt{2}\Delta^2\over 8|t|}$. Thus,
$E_0(6)$ has a weaker dependence on $\Delta$
than $E_0(7)$, similarly to the fully frustrated $t'=t$ cluster.

Our small cluster analysis indicates that the charge gap, $\Delta_c$,
increases with $\Delta$
due to the different natures of the two-electron
and three-electron bonds formed between
inequivalent sites. This result is not affected
by the presence of frustration in the cluster at the qualitative
level. However, geometrical
frustration ($t \sim t'$)
leads to qualitatively different
magnetic properties for different signs of $t$ in contrast
to the unfrustrated ($t'=0$) case.

\section{Ground state properties of the Ionic Hubbard model on a triangular lattice}
\label{sec:lanczos} Intermediate parameter regimes are explored
based on Lanczos diagonalization on finite clusters with N$_s$=12,
16 and 18 sites and periodic boundary conditions. Different cluster
shapes have been benchmarked against the exact solution of the
non-interacting model (\ref{ham}) and are shown in Fig.
\ref{figclusters}. The vectors defining the clusters are: ${\bf
T}_1=n_{11}{\bf a}_1 + n_{12} {\bf a}_2$ and ${\bf T}_2=n_{21} {\bf
a}_1 + n_{22}{\bf a}_2$, where $n_{1i}$ and $n_{2i}$ are integers. A
straightforward finite size scaling analysis is not possible because
of the complicated changes in the cluster shape as the lattice size
increases.

We  present results of the dependence of the charge order parameter, the charge gap and the spin
correlations on $\Delta$.
Numerical results are compared to the weak and strong coupling limits
as appropriate.

\subsection{Charge order}

The charge order parameter is first computed for $U=0$ and compared to
exact tight-binding results on the infinite lattice. This serves to calibrate the
importance of finite-size effects on a cluster. Second, the effect of $U$
on charge ordering is analyzed in detail.

The charge order parameter is
\begin{equation}
n_B-n_A=\sum_{{\bf k},\sigma} \langle  \Psi_0 |( c_{B{\bf k}\sigma}^\dagger  c_{B{\bf k}\sigma} -
c_{A{\bf k}\sigma}^\dagger c_{A{\bf k}\sigma} )|\Psi_0 \rangle,
\end{equation}
where $|\Psi_0\rangle$ is the ground state of the Hamiltonian (\ref{ham}).

\begin{figure}
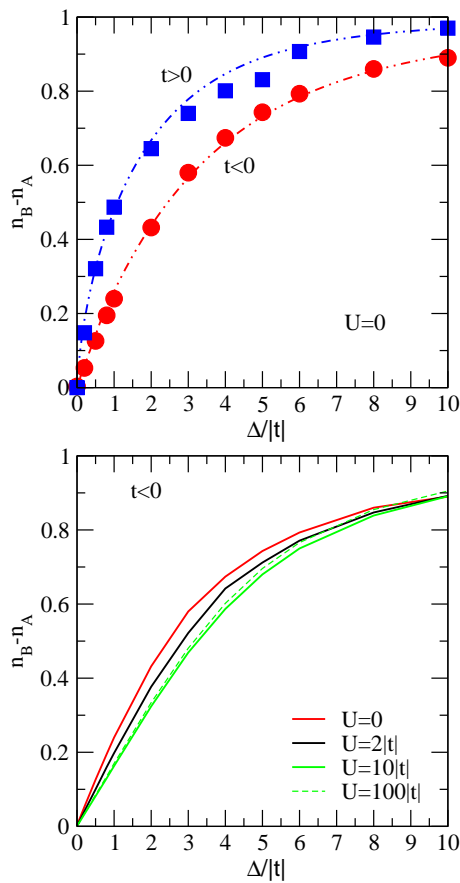

\begin{center}
\epsfig{file=fig7a.eps,width=6.cm,clip=}
\epsfig{file=fig7b.eps,width=6.cm,clip=}
\caption{\label{nAnBU}(Color online) Charge disproportionation,
 $n_B-n_A$, between inequivalent rows
in the ionic Hubbard model \ref{ham}.
In the top panel tight-binding (dash-dotted lines) exact results
for $U=0$ are compared with Lanczos diagonalization (open symbols) for the $N_s=18$
tilted cluster of Fig. \ref{figclusters} showing good agreement. In the
bottom panel the dependence
of $n_B-n_A$  with $U$ is shown from Lanczos diagonalization for $t<0$
on the same cluster.}
\end{center}
\end{figure}

In Fig. \ref{nAnBU} the charge order parameter is plotted for both signs of $t$ on
the $N_s$=18 site cluster of Fig. \ref{figclusters} and compared to the tight-binding result ($U=0$)
of the extended system. Since this cluster gives the
best agreement with the infinite limit of the
non-interacting model, among all of the clusters
that we studied, we mostly show results for this cluster in
the rest of this paper.

The effect of Coulomb repulsion on charge transfer is
also shown in Fig. \ref{nAnBU}. The qualitative
dependence of
charge transfer remains unchanged with $U$. However, increasing $U$ does
suppress $n_B-n_A$ a little for small and moderate $\Delta$.

In \Na the strong Coulomb interaction and weak charge transfer imply
\cite{MPM1,PMM}: $U>>\Delta$ and $\Delta \sim |t|$. Note also that
in this parameter regime charge transfer between A and B sites is
weak: $n_B-n_A<0.2$, for $\Delta \sim |t|$ which implies that a
charge transfer insulator formed by doubly occupied B sites
alternating with half-filled A sites is not possible.

\subsection{Reciprocal space charge ordering}

We now turn our attention to the charge populations of the one-electron (-) and (+)
hybrid bands obtained in Eq. (\ref{tb}) for $U=0$.
 The upper (lower) tight-binding band is
half-filled (filled) for any non-zero
$\Delta$ in the $t>0$ case \cite{PMM} while for $t<0$ this is
only the case for: $\Delta > 0.68|t|$. Simple arguments might then
suggest that if $U$ is sufficiently large then
the half-filled band may undergo a
Mott insulator transition.
 However, for such a single band argument
to be valid the half-filled + band must
be sufficiently high in energy above the filled-band so that
interband transitions induced by $U$ can be safely neglected.

In the basis of the non-interacting +,- band states the Hamiltonian is
\begin{widetext}
\begin{equation}
H=\sum_{{\bf k}\alpha,\sigma}  \epsilon_{\alpha}({\bf k}) c^+_{{\bf k}\alpha, \sigma}  c_{{\bf k}\alpha, \sigma}
\nonumber \\
+{1 \over N_s}\sum_{{\bf k, k', q}\alpha_1,\alpha_2,\alpha_3,\alpha_4} V
({\bf k -q} \alpha_1, {\bf k}\alpha_2, {\bf k' +q} \alpha_3, {\bf k' }\alpha_4)  c^+_{{\bf k-q}\alpha_1,\sigma_1}
  c^+_{{\bf k'+q}\alpha_3,\sigma_2} c_{{\bf k'}\alpha_4,\sigma_2} c_{{\bf k}\alpha_2,\sigma_1},
\nonumber \\
\end{equation}
\end{widetext}
where the
$\alpha$'s refer to the two values $+$ and $-$ and $V$ is the Coulomb matrix describing the 16
different scattering processes between the bands: $\epsilon_\alpha({\bf k})$ of Eq. (\ref{tb}).
The Coulomb matrix is
\begin{eqnarray}
V({\bf k_1}\alpha_1,{\bf k_2}\alpha_2,{\bf k_3}\alpha_3,{\bf k_4}\alpha_4) &=&
%\nonumber \\
\\
&& \hspace{-2cm} U\beta({\bf k_1}\alpha_1) \beta^*({\bf
k_2}\alpha_2)\beta({\bf k_3}\alpha_3)\beta^*({\bf k_4}\alpha_4),
\nonumber %\\
\end{eqnarray}
with
\begin{equation}
\beta({\bf k}\pm)={1 \over \alpha_{\mp{\bf k}}} { A_{{\bf k}\pm}
\over A_{{\bf k}\pm}-A_{{\bf k}\mp} },
\end{equation}
where $A_{{\bf k}\pm}$ is given by (\ref{mixing}) and $\alpha_{\bf k}$,
the normalization constant of Eq. (\ref{norm}).

The occupation of the non-interacting bands is obtained through the
expression: $n_{\pm}=\sum_{{\bf k} \sigma} \langle \Psi_0 |
c^+_{{\bf k}\pm\sigma}c_{{\bf k}\pm\sigma} |\Psi_0 \rangle$.

\begin{figure}
\begin{center}
\epsfig{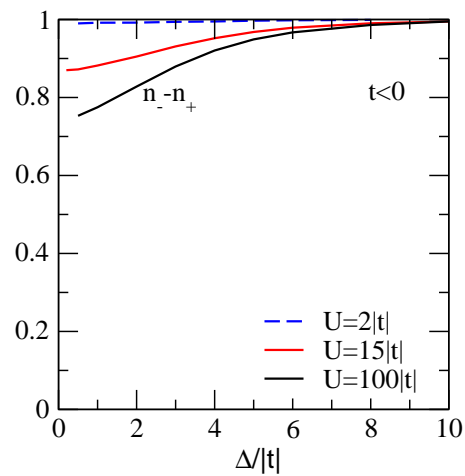}
\caption{\label{fignpnm}(Color online) Difference in filling of
the hybrid ${\pm}$-bands
as a function of $\Delta/|t|$ for $t<0$ and several values
of $U/|t|$.
Results are from Lanczos
calculations on 18-site clusters.
}
\end{center}
\end{figure}

In order to investigate the populations of the non-interacting bands
at large $U$ we plot
Lanczos results for $n_--n_+$ for $N_s=18$ in Fig. \ref{fignpnm}.
For $U=2|t|$, we find that $n_--n_+=1$.
However,  $n_--n_+<1$ for large $U$.
The interpretation of this result is complicated as $n_--n_+$ conflates two effects: (i) charge transfer between the bands,  effectively doping the $+$-band with electrons from the lower band; and (ii) for $U\ne0$ the bands are no longer eigenstates, thus the physical interpretation of $n_--n_+$ is unclear for large $U$. 
In spite of this interpretative difficulties it is interesting to note that the behavior seen in Fig. \ref{fignpnm}
differs from a recent mean-field approach\cite{PMM} which
includes local electron correlations only. However, at present, it is not possible to conclusively determine whether this is because non-local electron correlations may play an important
 role or because of the strong interband scattering induced by the
large $U$ which will eventually destroy the reciprocal space
description. This question is particularly important given the proposed 
role of tiny hole densities in the $+$-band\cite{PMM} in explaining the apparent discrepancy between the insulating behaviour suggested by resitivity,\cite{Foo,Balicas} ARPES,\cite{qian} and optical conductivity\cite{wang} and the observation of metallic quasiparticles via Shubnikov-de Haas experiments\cite{Balicas} on \Nan. This question requires future investigation.

\subsection{Charge gap}

In order to understand the electronic properties of model (\ref{ham}) we now discuss the
the charge gap and its dependence with charge order driven by $\Delta$ and
the Coulomb repulsion.
 The charge gap is defined by equation (\ref{chargegap0}).
When $t$ is small, the lowest order correction to excitation energies come
from the kinetic energy gain due to the propagation
along the B(A)-chains of a hole
(doublon) when extracting
(adding) an electron
to the zeroth order ground state configuration. Using degenerate perturbation
theory on Hamiltonian (\ref{tJ}) the gap to O($t^2/\Delta$) is
\begin{eqnarray}
\Delta_c &=& \Delta-2|t|+8t^2/\Delta-2t_{eff}^{(2)}-8t^3/\Delta^2
\nonumber  \\
&+&\delta E_{1D}^{t-J},
\label{chargegap}
\end{eqnarray}
where $\delta E_{1D}^{t-J}$ is the energy change
when adding a hole to a single half-filled A-chain modelled by the
$t-J$ model with $J$ given by
Eq. ({\ref{superex}). The second order hopping term $t_{eff}^{(2)}=2t^2/\Delta$,
contributes to the propagation of an antiholon in the A-chain. In the
antiferromagnetic case, $J>0$,
 the energy, $\delta E_{1D}^{t-J}$ in (\ref{chargegap}) per
A-chain site is given by the Bethe ansatz expression:\cite{ogata2}
\begin{eqnarray}
\delta E_{1D}^{t-J}&=& {-2|t|\over \pi}\sin(n_A\pi)
\nonumber \\
&-&8J(\ln2)n_A^2 \left[ {1-\sin(2n_A\pi) \over 2 n_A \pi } \right]
+J\ln2,
\end{eqnarray}
where $n_A=1-1/N^A_C$ is the number of electrons in a single A-chain of
$N^A_C$ sites when a single hole has been added to the otherwise half-filled
chain. In the ferromagnetic case ($J<0$)
\begin{equation}
E_{1D}^{t-J}=-2|t|-J/2.
\end{equation}
Hence, the charge gap is found to be larger for $t<0$ than $t>0$,
as shown in Fig. \ref{gap}, due to the geometrical frustration.
In contrast,
on the square lattice, $\Delta_c$, does not depend on the sign of $t$.
The dependence on the sign of $t$ becomes even more apparent for $\Delta \sim
|t|$, where
it is clear that for $t<0$ the gap is significantly larger than
for $t>0$.
Indeed, it may be that for small $\Delta/|t|$ and
 $t<0$ the system is an insulator and
for $t>0$ it is metallic.
However, finite size effects prevent us
from making a definitive statement about the existence
of a metallic state for small $\Delta/|t|$.
We note that dependence on the sign of $t$ is the opposite from
what one would expect from weak-coupling arguments.
For $t>0$ and $U=\Delta=0$ and at 3/4 filling the Fermi surface has
perfect nesting and there is a van Hove singularity in the
density of states at the Fermi energy.
Hence, weak-coupling arguments would suggest that for this sign of
$t$ the system would have a greater tendency to density wave instabilities
and insulating states. Further, our mean-field slave boson calculations\cite{PMM} also predict that the insulating state is more stable for $t>0$ than $t<0$, in contrast to the results reported in Fig. \ref{gap}. 

\begin{figure}
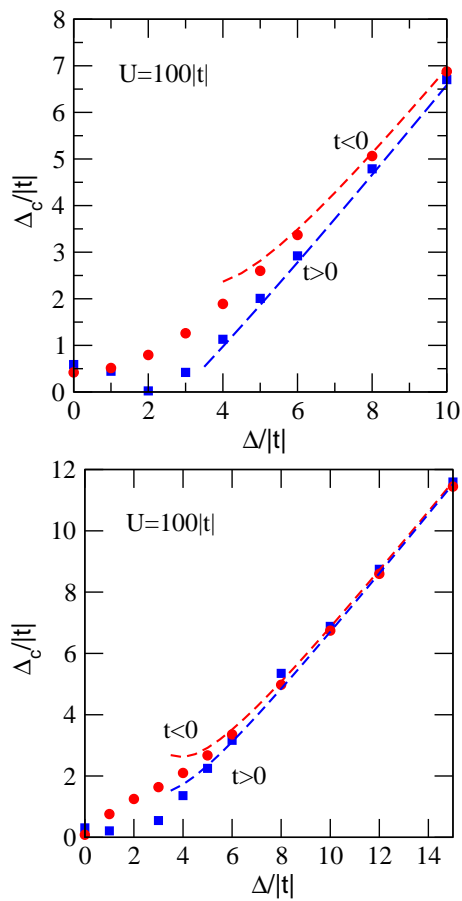

\begin{center}
\epsfig{file=fig9a.eps,width=6.0cm,clip=}\\
\epsfig{file=fig9b.eps, width=6.0cm, clip=}\\
\caption{\label{gap}(Color online) Dependence of the charge gap on   the
on-site potential $\Delta$.
The charge gap for $U=100|t|$ on a $N_s=18$ tilted cluster (top) and
a $N_s=16$ ladder-type cluster (bottom) (see  Fig. \ref{figclusters})
is shown. Dashed lines denote the results of the strong
coupling expansion (\ref{chargegap}) for comparison.
}
\end{center}
\end{figure}

%In Fig. \ref{chargeNs} the charge gap is plotted for different
%sizes in order to calibrate finite-size effects comparing
%$N_s=12, 16$ and 18 clusters for different values of $\Delta$ and
%$U$. The gap is found to be larger for larger values of $U$.
%Even for small values of $\Delta$ down to $\Delta=2|t|$ the gap
%is found to be finite.

%\begin{figure}
%\begin{center}
%\epsfig{file=gapU15Nstneg.eps,width=6.0cm,clip=}\\
%\epsfig{file=gapU100Nstneg.eps,width=6.0cm,clip=}\\
%\caption{\label{chargeNs}(Color online) Dependence of charge gap with size
%of the cluster. The charge gap for $t<0$, different $\Delta$ and $U$
%is plotted for tilted clusters of $N_s=12, 16$ and 18 sites of the type
%shown in Fig. \ref{figclusters}.
%{\bf The fact that the results depend on the shape of the cluster
%raises questions about the validity of this finite size scaling
%analysis. Maybe we should leave these figures out.}
%}
%\end{center}
%\end{figure}

%Summarizing, in the parameter range: $U>>\Delta$ and $\Delta \lesssim |t|$,
%relevant to \Nan we find an insulating state with small charge transfer
% $n_B-n_A<0.2$ and $n_--n_ \sim 0.7-0.8$. In this situation,
%a Mott insulator occurs
% for the $+$ band and the            $-$ band can be
%ignored because it is full.
% A CTI is not possible for such a weak charge transfer.  Deviations
%from the non-interacting band filling induced by $U$
%suggest a different insulating state of the covalent type (CI) with the
%hybridization between bands enhanced by the Coulomb repulsion.

\subsection{Magnetic order}
Spin correlations in the model are analyzed through
the static spin structure factor:
\begin{equation}
S({\bf q} )= {1 \over N_s} \sum_{ij} \exp^{i {\bf q} \cdot ({\bf R_i-R_j})}
<S^z_i S^z_j>
\end{equation}
where $S^z_j=(n_{j\uparrow}-n_{j\downarrow})/2$
is the $z$-component of the spin at the lattice site ${\bf R}_j$.

The dependence on $\Delta$ of the spin structure factor, $S({\bf q})$, is shown in Fig.
\ref{spin} for two different wavevectors with $U=100|t|$. For $t<0$ the results
 indicate a transition to the magnetically ordered state with wavevector
${\bf Q}_1=(\pi, \pi/\sqrt 3)$ (which
implies that the spins within the A-chains
are antiferromagnetically  ordered
and antiferromagnetically coupled with neighbouring A-chains) at about
$\Delta=2|t|$. This is the spin pattern shown in Fig. \ref{figstripes}.
For $t>0$ there is a range of $\Delta$ for which the A spins
are ferromagnetically coupled.
\begin{figure}
\begin{center}
\epsfig{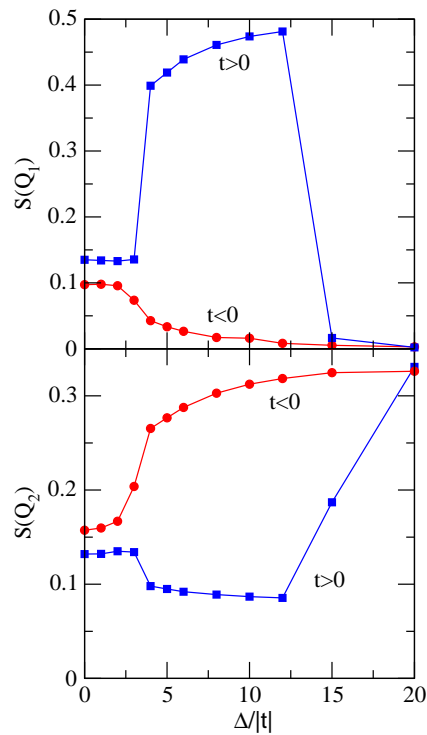}\\
\end{center}
\caption{\label{spin}(Color online)
Development of magnetic order.
 The static spin structure factor $S(\vec q)$ at
 the wavevectors
 (a) ${\bf Q}_1= (0,\pi/\sqrt{3}) $,
associated with C-type antiferromagnetism\cite{type-definition} (see Fig. \ref{figstripes},
left panel),
and
 (b) ${\bf Q}_2= (\pi,\pi/\sqrt{3}) $, associated with
G-type antiferromagnetism\cite{type-definition} (see Fig. \ref{figstripes},
right panel)),
 are shown for $U=100|t|$ as a function of
$\Delta$, the difference between the local energies on the A and B sublattices.
Calculations are performed on the $N_s=18$ cluster.
}
\end{figure}
The FM region obtained from Lanczos diagonalization is in agreement
with the condition $J<0$ extracted from a strong coupling expansion. This condition
gives a FM region for $5t< \Delta < \sqrt{2U}$.

The wavevector of
the magnetic order and the associated magnetic moment
of the ground state of the model with $t<0$ and $U \gg \Delta \gg |t|$
 (see Fig. \ref{figstripes})
are consistent with the antiferromagnetic ordering wavevector
observed in \Nan\cite{gasparovic,watanabe}. However, the observed
charge transfer between A and B chains in
\Nan\cite{bobroff,williams,argyriou} is much smaller than the
complete charge transfer sketched in Fig. \ref{figstripes}. Such a
large magnetic moment in the presence of a weak charge transfer
between A and B sites is not expected from either classical or weak
coupling arguments. Nevertheless, our previous exact diagonalization
calculations for $U>W$ (see Fig. 4 of Ref. [\onlinecite{MPM1}])
found a substantial magnetic moment and small charge transfer, even
for small $\Delta$, consistent with the experimental results.
% However, since the model effectively consists
%on a nearly half-filled hybrid band, strong electronic correlation effects are expected to occur.

\section{Dynamical properties}
\label{sec:dynamic}

In this section we discuss dynamical properties of the model (\ref{ham}).
We use Lanczos diagonalization on $N_s=18$ clusters to compute the
one-electron spectral density and the frequency dependent conductivity.

\subsection{One-electron spectral density}

The spectral density per spin is
\begin{eqnarray}
 A(\omega)&=&
\nonumber \\&& \hspace{-1cm} \sum_m|\langle \Psi_m(N-1)|
c_{i\sigma}|\Psi_0 \rangle|^2 \delta(\omega+(E_m(N-1)-E_0(N)))
\nonumber \\
&&\hspace{-1cm}+|\langle \Psi_m(N+1)| c^+_{i\sigma}|\Psi_0
\rangle|^2 \delta(\omega-(E_m(N+1)-E_0(N))),
\nonumber \\
\end{eqnarray}
where $E_m(N\pm1)$ is the spectra of excitations of the full quantum many-body
problem with $N\pm1$ electrons and $|\Psi_m(N\pm1)\rangle$ its associated wavefunctions.
$E_0(N)$ is the ground state of the
$N$ electron system with wavefunction $|\Psi_0 \rangle$.

In Fig. \ref{figawU100} we show the spectral density for $U=100|t|$
and $\Delta=10|t|$. As the system is well into the strong coupling
regime, $U >> \Delta >> |t|$, we can understand the main excitations
observed on the basis of the atomic limit ($t=0$).

\begin{figure}
\begin{center}
\epsfig{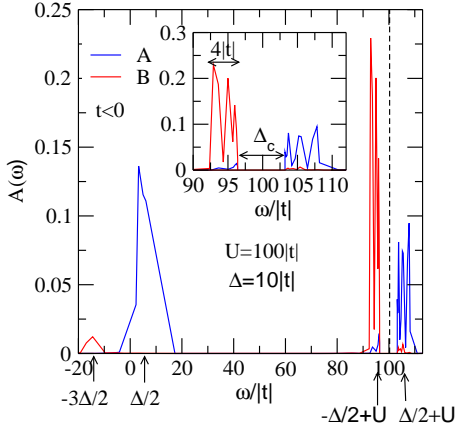}
\caption{\label{figawU100}(Color online) Density of states
of the charge transfer insulator. We take $U=100|t|$, $\Delta=10|t|$ and  $t<0$. The
inset shows the low energy part where the charge gap agrees with the
strong coupling expression Equation
(\ref{chargegap}).
The excitation energies in the atomic limit ($t=0$) are shown
by the arrows below the abcissa.
The chemical potential ($\omega=\mu$) is shown by the vertical dashed line.
}
\end{center}
\end{figure}

\begin{figure}
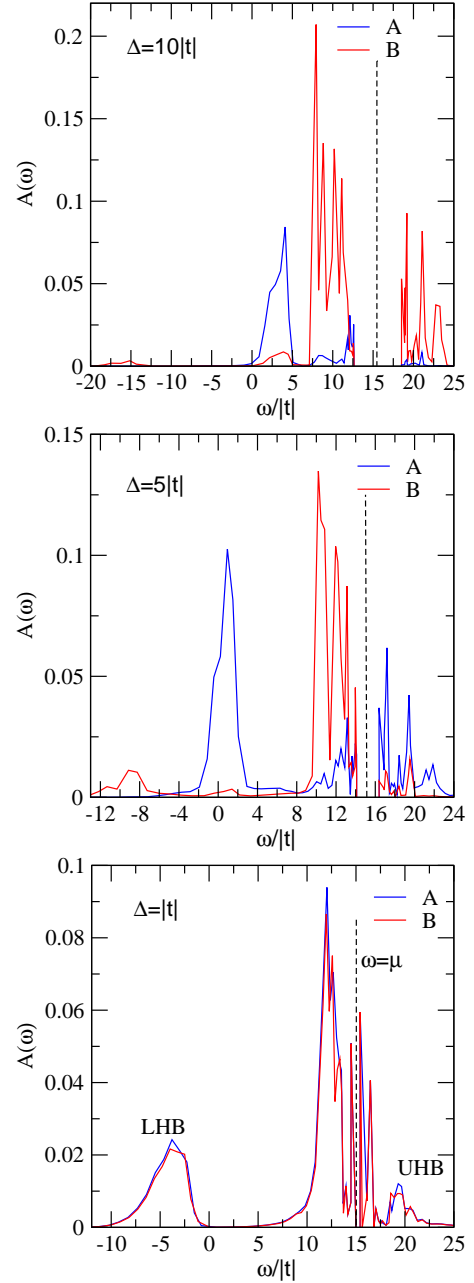

\begin{center}
\epsfig{file=fig12a.eps,width=6.0cm,clip=}
\epsfig{file=fig12b.eps,width=6.0cm,clip=}
\epsfig{file=fig12c.eps,width=6.0cm,clip=}
\caption{\label{figawU15}(Color online) Evolution of the
energy dependence of the density of states
with decreasing charge transfer.
  We take $U=15|t|$ and vary $\Delta$
from $10|t|$ (top, charge transfer
insulator) to
$\Delta=|t|$ (bottom, covalent insulator), all with $t<0$.
The chemical potential ($\omega=\mu$) is shown by the vertical dashed lines.
}
\end{center}
\end{figure}

For $t=0$ extracting an electron from the lattice can lead to three
possible excitation energies: $-3 \Delta/2$,  $\Delta/2$ and $
-\Delta/2+U$. While adding an electron we only have one excitation
energy at $\Delta/2+U$. Therefore, four peaks at these excitation
energies are expected in $A(\omega)$. A gap of $\Delta_c=\Delta$ is
therefore obtained when $t=0$ typical of a charge transfer
insulator. When the hopping is turned on, the lowest order
correction to excitation energies comes from the propagation of a
hole (doublon) along the B(A)-chains when extracting (adding) an
electron to the zeroth order ground state configuration. As the hole
(electron) added can be on any site of the chains the ground state
is $N_A$($N_B$)-fold degenerate. This degeneracy is lifted at the
first order in $t$, where the excitation energies for removing an
electron are $\Delta/2-2|t|(\sin(k)-1)$ and $\Delta/2-2|t|\cos(k)$,
from the A and B-sites respectively. Adding an electron to the
A-chain leads to a doublon with  excitation energy
$\Delta/2+U+2|t|(\sin(k)-1)$ Thus, a characteristic one-dimensional
broadening of $4|t|$ to the four peaks should be expected and the
gap is reduced from the atomic limit result to
$\Delta_c=\Delta-4|t|$. The hybridization between the chains lowers
the ground state energy for the $N$ electron configuration, due to
virtual excursions from a B-site to a nearest-neighbour A-site, by
$-4|t|^2/\Delta$ with no cost in Coulomb repulsion energy.
Therefore, the lowest $N+1$ electron excitation energy is pushed
upwards by $+4|t|^2/\Delta$ and the $N-1$ downward by
$-4|t|^2/\Delta$. This leads to an increase of the gap:
$\Delta_c=\Delta-4|t|+8t^2/\Delta$. The final charge gap, $\Delta_c$, including
the higher order corrections of Eq. (\ref{chargegap}) coincides with  
the numerical calculation shown in the inset of Fig. \ref{figawU100}.

In Fig. \ref{figawU15} we show the evolution of $A(\omega)$ with $\Delta$ for $U=15|t|$.
The four peak structure discussed above for the CTI remains for this smaller value
of $U$ and $\Delta=10|t|$.
As $\Delta$ decreases the peaks broaden
due to hybridization between the A and B chains and shift in energy.  For $\Delta=|t|$,
$A(\omega)$ contains a lower Hubbard band (LHB), an upper Hubbard band (UHB) and most
of the spectral weight is around the chemical potential $\omega=\mu$. We find that the energy
difference between the LHB and UHB is much larger than $U$, which we attribute to hybridization
 between the chains. Thus we identify this regime as a covalent insulator (CI).\cite{MPM1,sarma}

\subsection{Frequency dependent conductivity}

The incoherent part of the optical conductivity is calculated through the
current correlation function:
\begin{equation}
\sigma(\omega)= {\pi e^2 \over N_s} \sum_{m \neq 0}
{|\langle \Psi_m(N)|j_x| \Psi_0 \rangle|^2 \over E_m-E_0} \delta(\omega-(E_m-E_0)),
\end{equation}
where, $j_x$ is the $x$-component of the current operator, ${\bf
j}=it\sum_{<ij>,\gamma,\sigma} ({\bf R}_i-{\bf R}_j)
c^+_{i\sigma}c_{i+{\bf\gamma}\sigma}$, and  ${\bf R}_i$ is the
position of the $i^\textrm{th}$ lattice site. 
All nearest-neighbor sites entering the sum in the current 
are denoted by $\gamma$.

The evolution of the optical conductivity, $\sigma(\omega)$, with $\Delta$ is
shown in Fig. \ref{figoptU15}.  There are two main absorption bands in the optical spectra.
One is fixed at large energies of about $U$
and is associated with excitations between the Hubbard bands and a lower band which
shifts with $\Delta$. The lower absorption band is due to excitations associated with
transferring an electron from the $-$ band to the $+$ band. These produce a continuum of particle-hole
excitations of width of order $W \approx 9|t|$.
%Sharp features are found
%at the low energy edge of the $\Delta$ band and has a width of about $W=9|t|$.

We have also calculated the Drude weight for a range of parameters
in the model Hamiltonian. We do not show the results here because
due to finite size effects the detailed interpretation is not clear.
However, the trend is clear: as $U/|t|$ and $\Delta/|t|$
increase, the Drude weight decreases significantly.
\begin{figure}
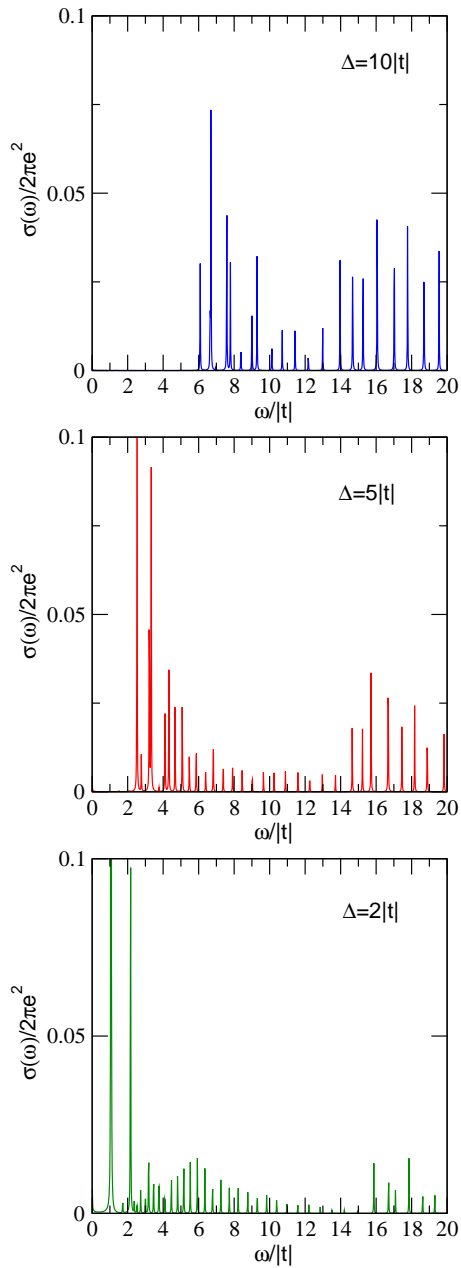

\begin{center}
\epsfig{file= fig13a.eps,width=6.0cm,clip=}
\epsfig{file= fig13b.eps,width=6.0cm,clip=}
\epsfig{file= fig13c.eps,width=6.0cm,clip=}
\caption{\label{figoptU15}(Color online) Frequency dependence of the optical
conductivity for several values of $\Delta/t$ and fixed $U=15|t|$ with $t<0$.
}
\end{center}
\end{figure}

\section{Conclusions}
\label{sec:concl}

We have considered the electronic properties of an ionic Hubbard model at $3/4$-filling
with stripes of alternating on-site potential.
 This model has a rich  phase diagram,
 in which various types of insulating and metallic states compete.
A charge transfer-type
insulator,  a Mott insulator and a covalent insulator occur in different
$U$-$\Delta$-$t$ parameter regimes.
The geometrical frustration of the triangular lattice leads to different
magnetic properties depending on the sign of $t$. For $t<0$ an antiferromagnetic interaction  occurs
whereas for $t>0$ a ferromagnetic coupling occurs in a broad range of parameters.

At $U>>\Delta>>|t|$, a
charge transfer insulator of doubly occupied chains of $B$-sites alternating with singly occupied
$A$-chains occurs, {\it i.e.}, $n_B-n_A=1$.
As $\Delta$ is decreased the system  remains insulating
although the charge disproportionation
between sites is incomplete: $n_B-n_A < 1$.

%Recent slave boson approaches to the same model \cite{PMM} with $U=\infty$ lead to a metal-to-insulator
%transition at a critical $\Delta_c$. The insulator, however, corresponds to
%the real space CTI insulator rather than to the hybridized one found here with exact diagonalization.
%This difference may be attributed to non-local electron
%correlations not included in the slave boson approach and deserves further investigation.

The insulating state of \Na is characterized by a small charge modulation,
a small charge gap, and strong Coulomb interaction.
Electronic structure calculations suggest that
\Na is in the parameter regime, $U>>|t|$ and
$\Delta \sim |t|$ which is different to the
strong coupling parameter regime. This regime
is difficult to analyze within weak coupling perturbation theory
 and numerical approaches
are helpful.
The model with no charge modulation, $\Delta=0$,
 reduces to a highly-doped Hubbard (or $t-J$)
model on a triangular lattice which is believed to be metallic.
However, under a weak external periodic potential, $\Delta \sim O(|t|)$,
our exact diagonalization analysis suggests a non-zero charge gap. Unlike
the charge transfer  insulators proposed previously
for \Na this
insulator is characterized by small real space charge transfer: $n_B-n_A << 1$.
% and non-zero interband charge transfer $n_--n_+<1$. As the + band deviates from
%half-filling a Mott insulator for this band is precluded. This filling deviation is induced
%by the effective strong Coulomb repulsion as for $U<<W$ the condition $n_--n_+=1$ is
An insulating state induced by strong hybridization of the
non-interacting bands
is realised and is reminiscent of the
covalent insulator (CI) description of some transition
metal oxides.\cite{sarma}

The behavior of the charge gap for $U>>|t|$ has been further explored by an exact analysis of two and four-site
clusters which indicates that the charge gap is always enhanced with $\Delta$ even
at small values.
A different dependence on $\Delta$ of the ground state many-body energies $E_0(N)$, $E_0(N-1)$
and $E_0(N+1)$ is found which reflects the different nature of the bonds formed between inequivalent
$A$ and $B$ sites. Two-electron bonds are well described as
 valence bonds which contain the effects
of strong electronic correlations whereas three-electron bonds are
accurately described by ``molecular'' orbitals which are
uncorrelated. While the former type of bond depends weakly on
$\Delta$ the latter does not.

%{\bf SHOULD WE LEAVE OUT THE FOLLOWING?}

%The magnetic pattern experimentally observed in neutron scattering experiments\cite{yokoi,gasparovic}
%finds a natural explanation within the present model for $t<0$
% and large $U$. % as shown in Fig. \ref{figstripes}.
%Substantial AF magnetic moments\cite{MPM1} are encountered in the model even for
%$\Delta \sim |t|$ when
%the charge disproportionation
% between A and B sites is small.
%These are attributed to the
%presence of strong correlation effects associated with the fact that hybridization
%effects still lead to a nearly half-filled band.

Optical conductivity experiments suggest a gap of the order of 0.020 eV ($\sim |t|/5$), a sharp
peak at about 0.026 eV ($\sim |t|/4$) which is at the lower edge of a continuum of
excitations which reaches energies up to about 0.9 eV ($\sim 9|t|$) \cite{wang}. This behavior is
consistent with the low energy adsorption band
found in the calculated $\sigma(\omega)$ (see Fig. \ref{figoptU15}) which is
located at about $\Delta \sim O(|t|)$ with
the continuum being the whole set of particle-hole excitations between the two hybridized bands which
spread over the whole bandwidth $\sim W$.

Recent experiments on \Nax  for $x=2/3$
the Na ions induce a charge ordering pattern with filled non-magnetic Co$^{3+}$ ions
arranged in a triangular lattice and Co$^{3.44+}$ magnetic sites forming a
kagom\'e lattice structure with the transferred holes moving on it. These
experiments are important as they relate the charge order with the different magnetic and
electronic properties of the material.  The present ionic Hubbard model
modified to include such ordering patterns could be used to explore the unconventional
metallic properties of \Nax at $x=2/3$.

We now briefly discuss the relationship between the results we
obtained here and those we recently obtained for the same model with
a slave boson mean-field theory.\cite{PMM}
 Slave bosons give an insulator only for $\Delta > 8|t|$
($\Delta > 5|t|$  ) for $t<0$ ($t<0$)
whereas exact diagonalisation suggests that the ground state
is insulating even for $\Delta \sim |t|$.

An important open question that this
study raises is,
what is the ground state for small $\Delta/t$?
The temperature dependence of the magnetic susceptibility of the
$t-J$ model on the triangular lattice has been calculated using exact
diagonalisation on small clusters.\cite{haerter}
The bottom left panel of Figure 6 in Ref. \onlinecite{haerter}
shows that for $J=0$ (i.e., $U \to \infty$ in the Hubbard model)
that at 3/4 filling that for all temperatures above about $0.4|t|$
that the susceptibility is the same as that for localised
 non-interacting spin-1/2 particles. The susceptibility has
a maximum at about $0.3|t|$ and then decreases with decreasing
temperature to a value about 2-3 times the value for $U=0$. These
results raise the question as to the nature of the ground state and
the tendency of the electrons to become localised and the spins to
antiferromagnetically order, even in the absence of an exchange
interaction, due to kinetic antiferromagnetism.\cite{shastry}

\appendix

\section{Heisenberg exchange couplings}
\label{sec:appI}

In this section we discuss the various contributions to the nearest
neighbour exchange coupling, $J$, between the $A$-sites. Taking
$|\Psi_0>$ as the ground state configuration in the strong coupling
limit $U>>\Delta>>t$, there is no correction to the lowest order in
the kinetic energy.  To O($t^2$)  we have the usual superexchange
antiferromagnetic contribution: $J=4t^2/U$. To O($t^3$) "ring"
exchange processes around the 3-site plaquette of the type shown in
Fig. \ref{ring3} remove the spin degeneracy and were already
discussed by Penc and collaborators  [\onlinecite{penc3}] in a
Hubbard model on a zig-zag ladder.  The energy of the singlet state
in Fig. \ref{ring3} is shifted by $4t^3/\Delta^2$ while the triplet
state by  $-4t^3/\Delta^2$. These shifts are opposite to the two
site case. As there are two possible ways of going around the
triangle in Fig \ref{ring3} and there are two neighbouring B-sites
(one below the two A-sites as shown in Fig. \ref{ring3} and another
above) the final contribution to the effective $J$ at O($t^3$) is
$J=E_t-E_s=-8t^3/\Delta^2$ enhancing the ferromagnetic tendencies as
compared to the ladder case \cite{penc3} by a factor of two.  In
contrast "ring" exchange processes around a 4-site plaquette
($O(t^4)$) of the type shown in Fig. \ref{ring4} lead to an AF
contribution to $J=40t^4/\Delta^3$. Including all possible exchange
processes, a total contibution to $J$ valid to ${\cal
O}(({t}/{\Delta})^4)$ is
\begin{equation}
J={4t^2 \over U}-{8t^3 \over \Delta^2}-{16t^3 \over  \Delta U}+{40t^4 \over \Delta^3} +{48t^4 \over \Delta^2 (2 \Delta +U)}+{16t^4 \over \Delta^2 U}
\end{equation}

\begin{figure}
\begin{center}
\epsfig{file=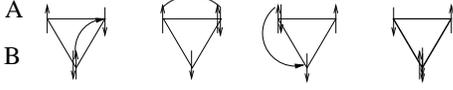,width=6.cm,clip=}
\caption{\label{ring3}(Color online)
Three site "ring" exchange processes contributing to the
exchange interaction, $J$,  between neighbour spins in an A chain of the
$t-J-J_\perp$ model (compare Equation (\ref{tJ})).}
\end{center}
\end{figure}

\begin{figure}
\begin{center}
\epsfig{file=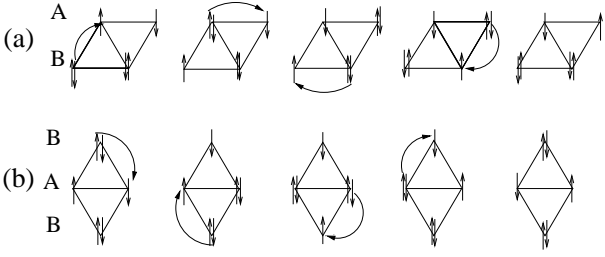,width=8.cm,clip=}
\caption{\label{ring4}(Color online)
Four site "ring" exchange processes of ${\cal O}(({t}/{\Delta})^4$ contributing to the
exchange interaction between two neighbouring sites in the $A$-chains in model (\ref{tJ}).}
\end{center}
\end{figure}

\section{Ground state wavefunctions for the four-site cluster and valence bond theory}
\label{sec:appII}

In this appendix, we discuss exact ground state wavefunctions
on the $N_s=4$ cluster of Fig. \ref{fig2sites} with $t'=t$ and $N=5$, 6 and 7 electrons.
Valence bond (VB) states,
 which are neutral configurations formed by two neighbour
 electrons in a singlet,\cite{pauling}
are found to describe the exact wavefunction accurately for $U>>|t|$
and $t<0$ for $N=6$. This is because ionic configurations have a
negligible weight in the full wavefunction at large-$U$ values.

\subsection{Ground state for $N=6$ electrons}

The exact ground state wavefunction, $|\Psi_0(6)>$, is well
described by the resonance between different possible VB states
between electrons inside the cluster:
\begin{widetext}
\begin{eqnarray}
|\Psi_0(6)> \approx a\left| \begin{array}{ccc}
  &\uparrow \downarrow &  \\
\color{red}{\uparrow} &    & \color{red}{\downarrow} \\
  &  \uparrow \downarrow &  \\
\end{array}\right>
+b\left| \begin{array}{ccc}
  &\color{red}{\uparrow} &  \\
\uparrow \downarrow &    & \uparrow \downarrow \\
  & \color{red}{\downarrow} &  \\
\end{array}\right>%\\
&+&c \left( \left| \begin{array}{ccc}
  & \color{red}{\downarrow} &  \\
\uparrow \downarrow &    & \color{red}{\uparrow} \\
  &    \uparrow \downarrow &  \\
\end{array}\right>
+\left| \begin{array}{ccc}
 & \color{red}{\downarrow} &  \\
\color{red}{\uparrow} &    & \uparrow \downarrow \\
  &    \uparrow \downarrow &  \\
\end{array}\right>
+ \left| \begin{array}{ccc}
  & \uparrow \downarrow &  \\
\uparrow \downarrow &    & \color{red}{\uparrow} \\
  &    \color{red}{\downarrow} &  \\
\end{array}\right>
+\left| \begin{array}{ccc}
 & \uparrow \downarrow &  \\
\color{red}{\downarrow} &    & \uparrow \downarrow \\
  &    \color{red}{\uparrow} &  \\
\end{array}\right>
 \right).
\label{rvb}
\end{eqnarray}
Here, and below, VB states are  colored  red. Note that the
horizontal sites are A sites while
 the vertical sites are B sites as
sketched in Fig. \ref{fig2sites}.
 For $U=\infty $ and $\Delta=0$, the energy of the
RVB state is $E^{RVB}_0(6)=-3.3723|t|$ and the weights of the
wavefunction are $a=b=0.4544$,  and $c=0.3831$. This RVB
wavefunction gives an accurate description of the exact ground state
energy which for $U=100|t|$ is: $E_0(6)=-3.453|t|$ and a
wavefunction described by (\ref{rvb}) with $a=0.459$, $b=0.445$,
$c=0.382$, plus small ionic terms. Note that, on the four site
cluster, the A and B sites are not equivalent even for $\Delta=0$
due to geometry of the cluster (cf. Fig. \ref{fig2sites})

The RVB wavefunction also accurately describes the
ground state energy in the limit: $U >> \Delta >> |t|$.
The  energy  of the RVB state
for $\Delta=10|t|$ is $E^{RVB}_0=-10.433|t|$, and the wavefunction
has weights
$a=0.977$, $b=0.021$, and $c=0.1058$. This is in good agreement
with the exact ground state, which has
 $E_0(6)=-10.453 |t|$ and $a=0.976$, $b=0.024$
and $c=0.108$ (plus small ionic terms). Thus the VB formed between
the two A sites dominates the wavefunction.

\subsection{ Ground state for $N=7$ electrons}

The 7-electron system contains one hole which
can hop around the cluster so the Coulomb interaction
has no effect. The ground state is simply a linear combination
of the states with one hole in the cluster:

\begin{eqnarray}
|\Psi_0(7)> \approx d\left(\left| \begin{array}{ccc}
  & \downarrow &  \\
\uparrow \downarrow &    & \uparrow \downarrow \\
  &    \uparrow \downarrow &  \\
\end{array}\right>+
\left| \begin{array}{ccc}
  & \uparrow \downarrow &  \\
\uparrow \downarrow &    & \uparrow \downarrow\\
  &    \downarrow &  \\
\end{array}\right> \right)
%\\
+e\left(\left| \begin{array}{ccc}
 & \uparrow \downarrow &  \\
\uparrow \downarrow &    & \downarrow \\
  &    \uparrow \downarrow &  \\
\end{array}\right>+
\left| \begin{array}{ccc}
 & \uparrow \downarrow &  \\
\downarrow &    & \uparrow \downarrow \\
  &    \uparrow \downarrow &  \\
\end{array}\right> \right).
\end{eqnarray}

%For $U=100|t|$ and $\Delta=10|t|$ the 7 electron ground
%state wavefunction is given by: $d=0.1227$  and
%$e=0.6964$. In the case $U=100|t|$ and $\Delta=0$, the ground
%state wavefunction is: $d=0.435$ and $e=0.557$.
Note that the $e$ and $d$ coefficients are different
even for $\Delta=0$ due to the geometry of the cluster, cf. Fig. \ref{fig2sites}.

\subsection{Ground state for $N=5$ electrons}

The $5$-electron ground state wave
function is well described by the resonance
between the following states:

\begin{eqnarray}
|\Psi_0(5)> \approx f\left(\left| \begin{array}{ccc}
  & \downarrow &  \\
\color{red}{\uparrow} &    & \color{red}{\downarrow} \\
  &    \uparrow \downarrow &  \\
\end{array}\right>+
\left| \begin{array}{ccc}
 & \uparrow \downarrow&  \\
\color{red}{\uparrow} &    & \color{red}{\downarrow} \\
  &  \downarrow &  \\
\end{array}\right> \right)
%\\
+g\left(\left| \begin{array}{ccc}
  & \color{red}{\uparrow} &  \\
\uparrow \downarrow &    & \downarrow\\
  & \color{red}{\downarrow} &  \\
\end{array}\right>+
\left| \begin{array}{ccc}
 & \color{red}{\uparrow} &  \\
\downarrow &    & \uparrow \downarrow \\
  &  \color{red}{\downarrow} &  \\
\end{array}\right> \right)
%\\
+h\left(\left| \begin{array}{ccc}
 & \downarrow &  \\
\uparrow \downarrow &    & \uparrow  \\
  &  \downarrow &  \\
\end{array}\right>
+\left| \begin{array}{ccc}
 & \downarrow &  \\
\uparrow &    & \uparrow \downarrow \\
 &  \downarrow &  \\
\end{array}\right> \right).
\end{eqnarray}
\end{widetext}

All many-body configurations contain VB singlets except for the last
term (proportional to $h$). The electrons in the A and B sites
prefer to align with antiferromagnetic order in contrast to
predictions of RVB.  This is due to the different number of spin up,
2, and down electrons, 3, for $N=5$. When a spin up is located at an
A site, the remaining two spin down electrons at the B sites
gain energy by aligining AF with the A electron. This is more
favorable energetically than having a VB betwen two electrons at A and B sites
and a down spin at the left B-site. 
For $U=100|t|$ and $\Delta=0$,
$f=0.423$, $g=0.326$ and $h=0.462$ with the ground state energy,
$E_0(5)=97.590|t|$.
For $\Delta=10|t|$ the wavefunction is approximately
given by $f=0.691$, $g=0.076$ and  $h=0.107$ with ground state
energy, $E_0(5)=94.615|t|$.

\acknowledgments We thank H. Alloul,
J. Bobroff, and R.R.P. Singh for helpful discussions.
J.M. acknowledges financial support
from MICINN (CTQ2008-06720-C02-02). B.J.P. was the recipient of an
Australian Research Council (ARC) Queen
Elizabeth II Fellowship (DP0878523). R.H.M. was the recipient  of
an ARC Professorial Fellowship (DP0877875).
This work was also supported by
the ARC Discovery Project Scheme (Projects DP0557532
and DP0878523). Some of the numerical calculations
were performed on the APAC national facility.

\end{document}